\documentclass[conference]{IEEEtran}

\usepackage{caption}
\usepackage{subcaption}
\usepackage{float}
\usepackage{verbatim}
\usepackage{listings}
\usepackage{algorithm}
\usepackage[noend]{algpseudocode}
\usepackage{color}
\usepackage[hidelinks]{hyperref}

\definecolor{dkgreen}{rgb}{0,0.6,0}
\definecolor{gray}{rgb}{0.5,0.5,0.5}
\definecolor{mauve}{rgb}{0.58,0,0.82}

\newcommand{\sectionname}{section}
\newcommand{\secref}[1]{\sectionname~\ref{#1}}
\newcommand{\figref}[1]{\figurename~\ref{#1}}
\newcommand{\lstref}[1]{Listing~\ref{#1}}
\newcommand{\numpy}{NumPy}
\newcommand{\sympy}{SymPy}
\newcommand{\Intel}{Intel\textsuperscript{\tiny{\textregistered}}}
\newcommand{\Xeon}{\Intel Xeon\texttrademark}
\newcommand{\XeonPhi}{\Intel Xeon Phi\texttrademark }

\usepackage{tikz}
\usepackage{pgfplots}
\usetikzlibrary{arrows,chains,positioning,fit,backgrounds,calc,shapes,
  shadows,scopes,decorations.markings,plotmarks}

\tikzstyle{line} = [draw, -, thick]
\tikzstyle{nodraw} = [draw, fill, circle, minimum width=0pt, inner sep=0pt]
\tikzstyle{box} = [line, rectangle, rounded corners, text centered]

\begin{document}

\title{
Devito: Towards a generic Finite Difference DSL using Symbolic Python
}

\author{
\IEEEauthorblockN{
Michael Lange\IEEEauthorrefmark{1},
Navjot Kukreja\IEEEauthorrefmark{2},
Mathias Louboutin\IEEEauthorrefmark{3},
Fabio Luporini\IEEEauthorrefmark{4},
Felippe Vieira\IEEEauthorrefmark{2},\\
Vincenzo Pandolfo\IEEEauthorrefmark{4},
Paulius Velesko\IEEEauthorrefmark{5},
Paulius Kazakas\IEEEauthorrefmark{6},
Gerard Gorman\IEEEauthorrefmark{1}}
\IEEEauthorblockA{\IEEEauthorrefmark{1}
Department of Earth Science and Engineering, Imperial College London, UK}
\IEEEauthorblockA{\IEEEauthorrefmark{2}
SENAI CIMATEC, Salvador, Brazil}
\IEEEauthorblockA{\IEEEauthorrefmark{3}
Seismic Lab. for Imaging and Modeling, The University of British Columbia, Canada}
\IEEEauthorblockA{\IEEEauthorrefmark{4}
Department of Computer Science, Imperial College London, UK}
\IEEEauthorblockA{\IEEEauthorrefmark{5}
College of Electrical and Computer Engineering, University of Oklahoma, USA}
\IEEEauthorblockA{\IEEEauthorrefmark{6}
Department of Computer Science, University of York, UK}
}

\maketitle

\begin{abstract}
Domain specific languages (DSL) have been used in a variety of fields
to express complex scientific problems in a concise manner and provide
automated performance optimization for a range of computational
architectures. As such DSLs provide a powerful mechanism to speed up
scientific Python computation that goes beyond traditional
vectorization and pre-compilation approaches, while allowing domain
scientists to build applications within the comforts of the Python
software ecosystem. In this paper we present Devito, a new finite
difference DSL that provides optimized stencil computation from
high-level problem specifications based on symbolic Python
expressions. We demonstrate Devito's symbolic API and performance
advantages over traditional Python acceleration methods before
highlighting its use in the scientific context of seismic inversion
problems.

\end{abstract}

\section{Introduction}

Python is one of the most popular high-level programming languages for
academics and scientists due to its clean syntax, great
expressiveness and a vast ecosystem of open source
packages~\cite{Oliphant2007,Millman2011,Perez2011}. Its direct use in
High Performance Computing (HPC), though, is complicated by the slow
execution within the Python interpreter itself. Several mechanisms
exist to overcome the interpreter overhead, ranging from static
compilation via Cython~\cite{Behnel2011} to generalised Just-In-Time
(JIT) compilation via Numba~\cite{Lam2015}.

However, for specific, often compute-intensive application areas such as
solving partial differential equations (PDE) with the finite element
method, domain specific languages (DSL) have also proven
successful~\cite{Logg2012}. DSLs aim to decouple the problem
specification from its low-level implementation to create a
separation of concerns between domain scientists and HPC specialists,
resulting in a direct payoff in productivity. Python is particularly
useful for creating DSLs, since its runtime type inference and
operator overloading via ``magic functions'' allows domain-specific
constructs to be easily embedded in the language itself.

Moreover, due to their restriction to a single problem domain,
code-generating DSLs can also augment the generated code to a
particular computational architecture, allowing hardware-specific
performance optimizations. In the age of specialized HPC
architectures, such as \XeonPhi or GPU accelerators, this not only
promises performance portability but also provides a valuable
alternative to the continuing search for new all-encompassing parallel
programming paradigms.

In this paper we present
Devito\footnote{\url{https://github.com/opesci/devito}}, a new finite
fifference DSL that utilises the \sympy{}~\cite{Meuer2016} package to
generate optimized parallel C code from high-level symbolic operator
definitions. Devito is primarily targeted at generating fast wave
propagation kernels for seismic inversion problems and therefore
offers an abstraction hierarchy that enables not only the generation
of fast stencil kernels, but also a range of domain-specific features,
such as sparse point interpolation, required in a real scientific
context. For this purpose Devito provides a two-level API that allows
users to define operator kernels as a mixture of high-level symbolic
equations and explicit C-like \sympy{} expressions to solve complex
inversion problems.

\section{Background}

\subsection{Speeding up Python computation}

Several methods exist to overcome the inherent interpreter overhead in
Python, most of which focus on using pre-compiled functions to perform
computationally intense loops. The most prominent example of this is
the \numpy{}~\cite{vanDerWalt2011} array interface, which manages
array data in contiguous blocks to avoid memory copies and provides
vectorized calculations, where operations are applied element-wise
over arrays using pre-compiled loops. The same technique is used by
the SciPy package that adds a wealth of pre-compiled mathematical
operators that can be applied in this vectorized way.

The effectiveness of \numpy{} vectorization, however, is limited by
the need to create arrays with matching data layout, which often
results in the ``outer loop'' being a pure Python loop. A more
aggressive pre-compilation can be achieved with
Cython~\cite{Behnel2011}, which allows users to annotate Python code
with additional type information and invoke an explicit compilation
step at install time at the expense of flexibility. Even more dynamic
compilation can be achieved with Numba~\cite{Lam2015}, a Just-In-Time
(JIT) compiler that focuses on array-oriented computation and utilises
LLVM to compile a subset of the Python language at runtime.

The Symbolic Python package \sympy{} used in Devito also provides
JIT compilation via the \lstinline!lambdify! utility function that
compiles a symbolic expression into either a Python function that
supports \numpy{} vectorization or pure C or Fortran kernels.

\subsection{DSLs embedded in Python}

Interest in building generic DSLs for solving PDEs is not new with
early attempts dating back as far as
1970~\cite{Cardenas1970,Grant1988,Yukio1985,VanEngelen1996}. More
recently, two prominent finite element software packages,
FEniCS~\cite{Logg2012} and Firedrake~\cite{Rathgeber2015}, have
demonstrated the power of symbolic computation using the DSL
paradigm. Both packages implement the Universal Form Language
(UFL)~\cite{Alnaes2014} that allows scientists to express complex
finite element problems symbolically in the weak form. Reducing the
complexity of a problem to its symbolic definition then enables
further symbolic abstraction, such as the automated generation of
adjoint models, as demonstrated by Dolfin-Adjoint~\cite{Farrell2013}.
It is noteworthy that FEniCS, Firedrake and Dolfin-Adjoint all provide
a native Python interface to the user, with the DSL embedded in the
Python language itself.

The optimization of regular grid and stencil computations has also
produced a vast range of libraries and DSLs that aim to ease the
efficient automated creation of high-performance
codes~\cite{Brandvik2010,Henretty2013,Zhang2012,Datta2008,PATUS,STARGATES,HiPACC}.
Most stencil DSLs, however, define their own custom input language,
which limits their scope and applicability for practical
applications. Since solving realistic problems often requires more
than just a fast and efficient PDE solver, scientific applications can
benefit from DSLs embedded in Python directly due to the native
compatibility with the Python software ecosystem.

\subsection{Abstraction layers in DSL design}
\label{sec:dsl-abstraction-layers}

While DSL-based approaches promise great flexibility and productivity,
care needs to be taken when designing such domain-specific software
stacks. In particular the choice of abstraction layers and interfaces
is crucial, since future development and practical use can often be
hampered by too narrow abstractions, while all-encompassing packages
often become too generic, complicated and hard to maintain. The
separation of the high-level symbolic problem definition layer from
the low-level implementation details of the generated code is of
crucial importance here, since it provides a separation of concerns
between the domain scientists and HPC specialists. This not only
increases productivity on both sides, but also ensures maintainability
of the generated software stack.

Active libraries, such as OP2~\cite{Mudalidge2012} for unstructured
mesh applications or OPS~\cite{Reguly2014} for structured grids, aim
to decouple the problem specification from the underlying execution by
providing an abstraction of data movement following the access-execute
paradigm~\cite{Howes2008}. A particular implementation of such an
abstraction is provided by PyOP2~\cite{Rathgeber2012}, which acts as
the underlying unstructured mesh traversal engine for the previously
mentioned Firedrake finite element system. Similar mechanisms
are provided by the Halide project~\cite{Halide}, a domain specific 
language, compiler, and auto-tuning system that was originally designed 
for image processing pipelines

\section{Devito design and API}
\label{sec:devito}

In contrast to traditional Python-embedded DSLs, Devito does not
define its own high-level language {\em per se}. Instead, it relies
on \sympy{} to provide the building blocks for users to express a
finite difference operator as a symbolic equation. Devito, however,
annotates these objects with additional meta-data and provides new
abstractions specific to defining finite difference problems. This
allows Devito to automate the transformation of high-level symbolic
expressions to optimized executable C code by maintaining and
manipulating \sympy{} expression objects throughout the code generation
process, while leveraging symbolic manipulation utilities provided by
\sympy{}.

The automated code generation process is performed in multiple steps
and follows the {\em Principle of Graceful Degradation}, which
suggests that users should be able to circumvent limits to the top
layer abstraction by utilising a lower-level API. A common example for
this is sparse point interpolation, which is required by many
scientific applications to extract solution values at specific points,
but can not easily be expressed as a high-level PDE. For such cases
Devito provides users with a lower level API to allow custom
expressions to be added to the generated code in a C-like manner.

\figref{fig:devito_layers} highlights the overall design hierarchy of
Devito and shows the individual layers of transformations. At the top,
user can define data object that associate \lstinline!sympy.Function!
symbols with data buffers and use them to derive stencil equations in
a concise symbolic format. The resulting stencil expressions are
automatically expanded and transformed into explicit array accesses
when the user builds a \lstinline!devito.Operator! object from the
\sympy{} expression. A secondary low-level kernel API is then utilised
to generate optimized C code from the intermediate representation, and
a set of compiler presets is used to perform Just-in-Time (JIT)
compilation before executing the kernel operator over the given data.

\tikzset{
>=stealth',
  concept/.style={
    box,
    text width=18em, 
    minimum height=3em, 
  },
  explanation/.style={
    xshift=6.5cm,
    text width=25em,
    minimum height=3em, 
    text centered
  },
  line/.style={
    draw, thick, <-
  },
  every join/.style={
    ->, thick,shorten >=1pt
  },
  decoration={brace},
  tuborg/.style={decorate},
  tubnode/.style={midway, right=2pt},
}
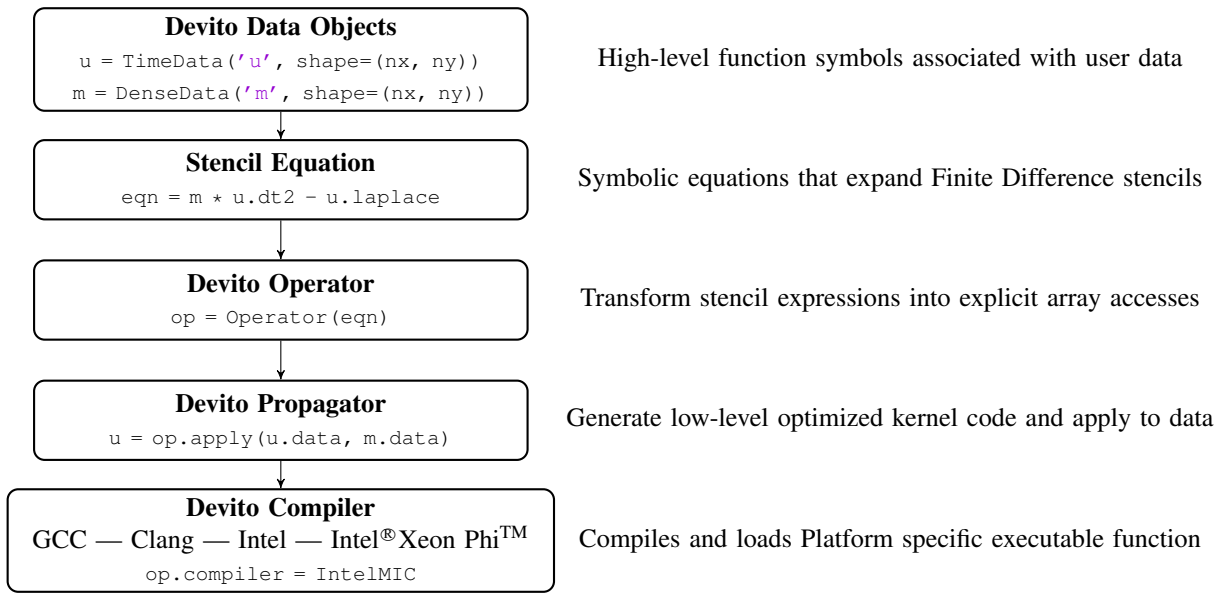
\begin{figure*}[ht]\centering
  \begin{tikzpicture} [node distance=1.6cm]
    \node[concept] (data) {
      {\bf Devito Data Objects}\\
      \lstinline!u = TimeData('u', shape=(nx, ny))!\\
      \lstinline!m = DenseData('m', shape=(nx, ny))!
    };
    \node[concept, below of=data] (eqn) {
      {\bf Stencil Equation}\\
      \lstinline!eqn = m * u.dt2 - u.laplace!
    };
    \node[concept, below of=eqn] (operator) {
      {\bf Devito Operator}\\
      \lstinline!op = Operator(eqn)!
    };
    \node[concept, below of=operator] (kernel) {
      {\bf Devito Propagator}\\
      \lstinline!u = op.apply(u.data, m.data)!
    };
    \node[concept, below of=kernel, text width=20em] (compiler) {
      {\bf Devito Compiler}\\
      GCC | Clang | Intel | \XeonPhi \\
      \lstinline!op.compiler = IntelMIC!
    };
    \node[explanation, right of=data] {
      High-level function symbols associated with user data
    };
    \node[explanation, right of=eqn] {
      Symbolic equations that expand Finite Difference stencils
    };
    \node[explanation, right of=operator] {
      Transform stencil expressions into explicit array accesses
    };
    \node[explanation, right of=kernel] {
      Generate low-level optimized kernel code and apply to data
    };
    \node[explanation, right of=compiler] {
      Compiles and loads Platform specific executable function
    };
    \draw[->] (data) edge (eqn) (eqn) edge (operator);
    \draw[->] (operator) edge (kernel) (kernel) edge (compiler);
  \end{tikzpicture}
  \caption{The Devito code generation process utilises multiple
    abstraction layers that allow users to customise each individual
    transformation in the toolchain, ranging from high-level symbolic
    equation definitions to low-level compiler settings.}
  \label{fig:devito_layers}
\end{figure*}

\subsection{High-level API: Derivatives and equations}
\label{sec:devito_high}

One of the core ideas behind Devito is to associate simulation data
with symbolic objects to leverage the symbolic manipulation
capabilities provided by \sympy{} during code generation. For this
purpose Devito provides \lstinline!DenseData! and \lstinline!TimeData!
objects that behave like \lstinline!sympy.Function! objects, but also
encapsulate data buffers and present them to the user as wrapped
\numpy{} arrays. Importantly, Devito is thus able to manage data
allocation internally to provide data caching on disk and first touch
allocation to ensure data locality on NUMA systems.

To ensure associated meta-data is not lost during symbolic
manipulation, Devito maintains its own symbol cache that
``re-attaches'' Devito-specific information whenever a new instance of
a symbol is created. Since \sympy{} uses frequent re-instantiation of
\lstinline!Function! objects to facilitate symbolic substitutions in
expressions, this enables Devito symbols to behave opaquely during
symbolic manipulation, for example when forming finite difference
stencils. The following example demonstrates how a two-dimensional
symbolic data object \lstinline!f(x, y)!  maintains the shape
information of its associated data buffer (\lstinline!f.data!), while
expanding the first derivative in dimension \lstinline!x! using
\sympy{}'s \lstinline!as_finite_diff! utility.

\begin{minipage}{\linewidth}
\begin{lstlisting}
In [1]: from Devito import DenseData
In [2]: from sympy import as_finite_diff
In [3]: from sympy.abc import x, h

In [4]: f = DenseData(name='f', shape=(10, 12))
Out[4]: f(x, y)

In [5]: eqn = as_finite_diff(f.diff(x), [x, x+h])
Out[5]: -f(x, y)/h + f(h + x, y)/h

In [6]: f_h = eqn.args[0].args[1]
Out[6]: f(h + x, y)

In [7]: f_h.data.shape
Out[7]: (10, 12)
\end{lstlisting}
\end{minipage}

Moreover, Devito objects also provide shorthand notation for common
finite difference formulations, such as first, second and cross
derivatives in the space and time dimensions. The order of the stencil
discretization is hereby defined on the symbolic object itself, which
allows us to express the second derivative of \lstinline!f! in
\lstinline!x! as:

\begin{minipage}{\linewidth}
\begin{lstlisting}
In [1]: from devito import DenseData

In [2]: f = DenseData(name='f', shape=(10, 10),
                      space_order=2)
Out[2]: f(x, y)

In [3]: f.dx2
Out[3]: -2*f(x, y)/h**2 + f(-h + x, y)/h**2
        + f(h + x, y)/h**2
\end{lstlisting}
\end{minipage}

In the resulting stencil expression Devito inserts the symbol
\lstinline!h! to denote grid spacing in the \lstinline!x! dimension.
It is important to note here that this notation allows users to
quickly increase the order of the stencil discretization by simply
changing a single parameter during symbol construction.

Another utility property of Devito's symbolic data objects is the
\lstinline!f.laplace! operator that expands to \lstinline!(f.dx2 + f.dy2)!
for two-dimensional problems and \lstinline!(f.dx2 + f.dy2 + f.dz)! for
three-dimensional ones. This allows the dimensionless expression of
complex equations, such as a simple wave equation, to be defined very
concisely:

\begin{minipage}{\linewidth}
\begin{lstlisting}
In [1]: from devito import DenseData
In [2]: from sympy import Eq

In [3]: u = TimeData(name='u', space_order=2,
                     time_order=2, shape=(10, 12))
In [4]: m = DenseData(name='m', space_order=2
                      shape=(10, 12)))

In [5]: eqn = Eq(m * u.dt2, u.laplace)
Out[5]: Eq((-2*u(t, x, y)/s**2
            + u(-s + t, x, y)/s**2
            + u(s + t, x, y)/s**2)*m(x, y),
            -4*u(t, x, y)/h**2
            + u(t, x, -h + y)/h**2
            + u(t, x, h + y)/h**2
            + u(t, -h + x, y)/h**2
            + u(t, h + x, y)/h**2)
\end{lstlisting}
\end{minipage}

The most important abstraction provided by Devito however is the
\lstinline!Operator! class that allows users to trigger the automated
code generation and compilation cascade. Users simply need to provide an
expanded stencil expression and optional compiler settings via
\lstinline!op = Operator(eqn, compiler=IntelCompiler)!. The resulting
operator object can then be used to execute the generated kernel code
via \lstinline!op.apply(u, m)!.

\subsection{Low-level API: Expressions and loops}
\label{sec:devito_low}

One of the key transformations performed during the code generation
process is to resolve the grid spacing variables, for example
\lstinline!h!, into explicit grid indices. During this transformation
Devito converts symbolic functions like \lstinline!f(x, y)! into
\lstinline!sympy.Indexed! expressions of the form \lstinline!f[x, y]!,
before converting the dimension symbols to loop counters to yield
\lstinline!f[i1, i2]!. It is important to note here that the
intermediate indexed representation with explicit dimensions is also
accepted input when creating \lstinline!Operator! instances, allowing
users to input custom expressions with irregular grid indexing. The
low-level API is also a useful tool for developers to rapidly test
and develop additional higher-level abstractions.

Moreover, a new low-level API is currently under development that
will allow the automatic inference of iteration spaces and loop
constructs from the indexed stencil expression itself. The key
to this are \lstinline!Dimension! objects that act like index symbols
(\lstinline!x!, \lstinline!y! or \lstinline!t!) and define the
iteration space over which to loop in the generated kernel code, while
defining additional meta-information like boundary padding layers or
buffer switching for time loops. This new API will allow the generic
and automated construction of arbitrary loop nests via provided
\lstinline!Iteration! and \lstinline!Expression! classes, enabling the
rapid development of further mathematical abstractions in the
high-level API. This will also provide the possibility of closer
integration with dedicated access-execute frameworks (see
\secref{sec:dsl-abstraction-layers}) in the future to further leverage
automated loop-level performance optimizations.

\subsection{Performance optimization}
\label{sec:devito_performance}

Devito provides a set of automated performance optimizations during
code generation that allow user applications to fully utilise the
target hardware without changing the model specification. The
optimizations provided by Devito range from symbolic manipulations
applied to the provided stencil expressions to hardware-specific
implementation choices like loop blocking.

\subsubsection{Shared-memory Parallelism}

Devito provides thread-parallel execution through automated code
annotation with OpenMP pragmas inserted at the appropriate locations
in the code, with a single thread pool allocated outside of the
primary timestepping loop. On multi-socket systems Devito performs so
called first touch memory allocation on its symbolic data objects
ensuring high data locality on NUMA architectures.

\subsubsection{Vectorization}

SIMD parallelism and vectorization are a crucial performance
optimization for compute intensive stencil codes to fully exploit
modern HPC architectures. Devito ensures efficient vectorization by
inserting toolchain-specific compiler hints into the generated code
and enforcing data alignment on page boundaries during allocation.

\subsubsection{Loop Blocking}

Loop or cache blocking is a well-known performance optimization for
stencil codes that aims to increase data locality and thus cache
utilisation by splitting iteration spaces (loops) into spatially
aligned blocks. The effectiveness of the technique is highly dependent
on the cache size of the target architecture. Devito therefore
provides an interfaces for users to define block sizes explicitly, as
well as an ``auto-tune'' mode where a defined \lstinline!Operator!
object is used to identify the optimal block size through a
brute-force search.

\subsubsection{Common sub-expression elimination}

On top of architecture-specific optimizations Devito also aims to
leverage symbolic optimizations to speed up computation. An important
example of this is common sub-expression elimination (CSE) that aims
to reduce the amount of redundant computation within complex stencil
expressions by factoring out common terms. While it may be argued that
most modern optimizing compilers are able to do this, it is important
to note here that Devito utilises \sympy{}'s internal CSE functionality
to apply the technique at a much higher abstraction level.

As a result, processing times of symbolic expressions are reduced
during the code generation phase, and the compilation time of the
generated code itself is improved. This is important since complex
stencils can increase compilation time to several hours if CSE is
performed by the compiler, in particular for high-order seismic
inversion kernels that often contain very large numbers of terms.

\section{2D Diffusion Example}

The use of symbolic notations to define finite difference operators
strongly enhances the efficiency of model developers and allows well
known problems to be expressed in a concise manner. In this section we
will first demonstrate the use of symbolic Python notation in comparison
to pure Python code on a simple two-dimensional diffusion example. We
will also analyse the performance benefits of Devito operators over
vectorized \numpy{} array notation, before demonstrating the
construction of seismic inversion operators using Devito's symbolic
interfaces in the next section.

The general diffusion or heat equation in two dimensions is commonly
notated as
\begin{equation} \label{eqn:diffusion}
  \frac{\partial u}{\partial t} = \alpha\nabla^2 u
\end{equation}
where $u$ denotes the unknown function to solve for, $\alpha$ is the
diffusion coefficient and $\nabla^2u$ denotes the Laplace operator, as
defined in \secref{sec:devito_high}. To keep the following examples
short we will be using a forward Euler approach with constant grid
spacing in $x$ and $y$, while the the timestep size is computed as
\begin{equation} \label{eqn:diffusion_dt}
\Delta t = \frac{\Delta x^2\Delta y^2}{2\alpha (\Delta x^2 + \Delta
  y^2)}
\end{equation}
where $\Delta x$, $\Delta y$ denotes the grid spacing and $\Delta t$
is the timestep size.

\subsection{Pure Python}

A first direct implementation of the described equation using an
alternating buffer scheme for \lstinline!u! in pure Python is shown in
\lstref{lst:diff_python}. The implementation consists of three loops
that apply a 5-point star stencil - a 3-point stencil in \lstinline!x!
(\lstinline!uxx!) and a 3-point stencil in \lstinline!y!
(\lstinline!uyy!) to compute the update. It is important to note here
that the stencil is only applied to elements in the interior of the
grid in order to avoid out-of-bound array accesses.

\begin{lstlisting}[
    caption= Diffusion code using pure Python loops.,
    label={lst:diff_python}, float, floatplacement=HT]
_, nx, ny = u.shape
for t in range(timesteps):
    t0 = t % 2
    t1 = (t + 1) % 2
    for i in range(1, nx-1):
        for j in range(1, ny-1):
            uxx = (u[t0, i+1, j] - 2*u[t0, i, j]
                   + u[t0, i-1, j]) / dx2
            uyy = (u[t0, i, j+1] - 2*u[t0, i, j]
                   + u[t0, i, j-1]) / dy2
            u[t1, i, j] = u[t0, i, j]
                          + dt * a * (uxx + uyy)
\end{lstlisting}

\subsection{Vectorized \numpy{} arrays}

Since the loop performance overhead of the Python interpreter is well
known, the most common optimization of simple arithmetic computations
on array and grid data is to utilise \numpy{} vectorization. This
applies arithmetic operations element-wise over all elements in an
array using pre-compiled kernels. For stencil expressions it is
therefore important to adjust the bounds of the array view that
represents each stencil point, as demonstrated in
\lstref{lst:diff_numpy}.

\begin{lstlisting}[
    caption=Diffusion code with vectorized \numpy{} arrays.,
    label={lst:diff_numpy}, float,floatplacement=HT]
for t in range(timesteps):
    t0 = t % 2
    t1 = (t + 1) % 2
    uxx = (u[t0, 2:, 1:-1] - 2*u[t0, 1:-1, 1:-1]
           + u[t0, :-2, 1:-1]) / dx2
    uyy = (u[t0, 1:-1, 2:] - 2*u[t0, 1:-1, 1:-1]
           + u[t0, 1:-1, :-2]) / dy2
    u[t1, 1:-1, 1:-1] = u[t0, 1:-1, 1:-1]
                        + a * dt * (uxx + uyy)
\end{lstlisting}

\subsection{Symbolic Python - lambdified}

An alternative approach to specifying the loop kernel that implements
a particular equation directly is to define it symbolically via
\sympy{}, as demonstrated in \lstref{lst:diff_sympy}. The diffusion
equation can thus be expressed in a manner that is mathematically 
concise, while \sympy{} utilities are used to expand the derivative 
stencils and create an executable function via \lstinline!sympy.lambdify!. 
It is noteworthy that the \lstinline!'numpy'! mode argument allows us to utilise
vectorized operations on sub-arrays, similar to the pure \numpy{}
variant.

\begin{lstlisting}[
    caption=Diffusion code using symbolic SymPy objects.,
    label={lst:diff_sympy}, float, floatplacement=HT]
from sympy import Eq, Function, lambdify
from sympy.abc import x, y, t, s, h
p = Function('p')
dx2 = as_finite_diff(p(x, y, t).diff(x, x),
                     [x - h, x, x + h])
dy2 = as_finite_diff(p(x, y, t).diff(y, y),
                     [y - h, y, y + h])
dt = as_finite_diff(p(x, y, t).diff(t), [t, t+s])
eqn = Eq(dt, a * (dx2 + dy2))
stencil = solve(eqn, p(x, y, t + s))[0]
subs = (p(x, y, t), p(x+h, y, t), p(x-h, y, t),
        p(x, y+h, t), p(x, y-h, t), s, h, a)
kernel = lambdify(subs, stencil, 'numpy')

for ti in range(timesteps):
    t0 = ti % 2
    t1 = (ti + 1) % 2
    u[t1, 1:-1, 1:-1] = kernel(
        u[t0, 1:-1, 1:-1], u[t0, 2:, 1:-1],
        u[t0, :-2, 1:-1], u[t0, 1:-1, 2:]
        u[t0, 1:-1, :-2], dt, dx)
\end{lstlisting}

\subsection{Symbolic Python - Devito}
\label{sec:diffusion_devito}
A similar symbolic approach is provided by Devito, where the expansion
of the derivative stencils is automated by the provided symbolic data
objects. As shown in \lstref{lst:diff_devito}, this allows the stencil
discretization to be adjusted via a single parameter change, while the
equation definition and reorganisation of the stencil expression via
\lstinline!sympy.solve! is identical to the pure \sympy{} approach.
The \lstinline!devito.Operator! created from the expression is then
used to generate low-level code and execute it for a given number of
timesteps.

\begin{lstlisting}[
    caption=Diffusion code using the symbolic Devito API.,
    label={lst:diff_devito}, float, floatplacement=HT]
from devito import TimeData, Operator
from sympy.abc import s, h
u = TimeData(name='u', shape=(nx, ny),
             time_order=1, space_order=2)
u.data[0, :] = ui[:]

eqn = Eq(u.dt, a * (u.dx2 + u.dy2))
stencil = solve(eqn, u.forward)[0]
op = Operator(stencils=Eq(u.forward, stencil),
              subs={h: dx, s: dt}, nt=timesteps)
op.apply()
\end{lstlisting}

An example of the auto-generated C code, is shown in
\figref{lst:diff_devito_c}. The example demonstrates Devito's use of
parallelisation and vectorization pragmas, as well as its internal
resolution of alternating buffer access. Moreover it can be seen that
Devito has replaced all numerical constants, such as the diffusion
coefficient \lstinline!a! and spacing parameters \lstinline!dx! and
\lstinline!dt!, and inserted all known loop dimensions to aid compiler
optimizations.

\begin{lstlisting}[
    basicstyle={\scriptsize\ttfamily},
    caption=Auto-generated C code to solve diffusion example.,
    label={lst:diff_devito_c}, float, floatplacement=HT]
extern ``C'' int Operator(float *u_vec)
{
  float (*u)[1000][1000] = (float (*)[1000][1000]) u_vec;
  {
    int t0;
    int t1;
    #pragma omp parallel
    for (int i3 = 0; i3<500; i3+=1)
    {
      #pragma omp single
      {
        t0 = (i3)%(2);
        t1 = (t0 + 1)%(2);
      }
      {
        #pragma omp for schedule(static)
        for (int i1 = 1; i1<999; i1++)
        {
          #pragma omp simd aligned(u:64)
          for (int i2 = 1; i2<999; i2++)
          {
            u[t1][i1][i2] = 2.5e-1F*u[t0][i1][i2-1]
                          + 2.5e-1F*u[t0][i1][i2+1]
                          + 2.5e-1F*u[t0][i1-1][i2]
                          + 2.5e-1F*u[t0][i1+1][i2];
          }
        }
      }
    }
  }
\end{lstlisting}

\subsection{Performance comparison}

\begin{figure}[ht] \centering
\includegraphics[width=.4\textwidth]{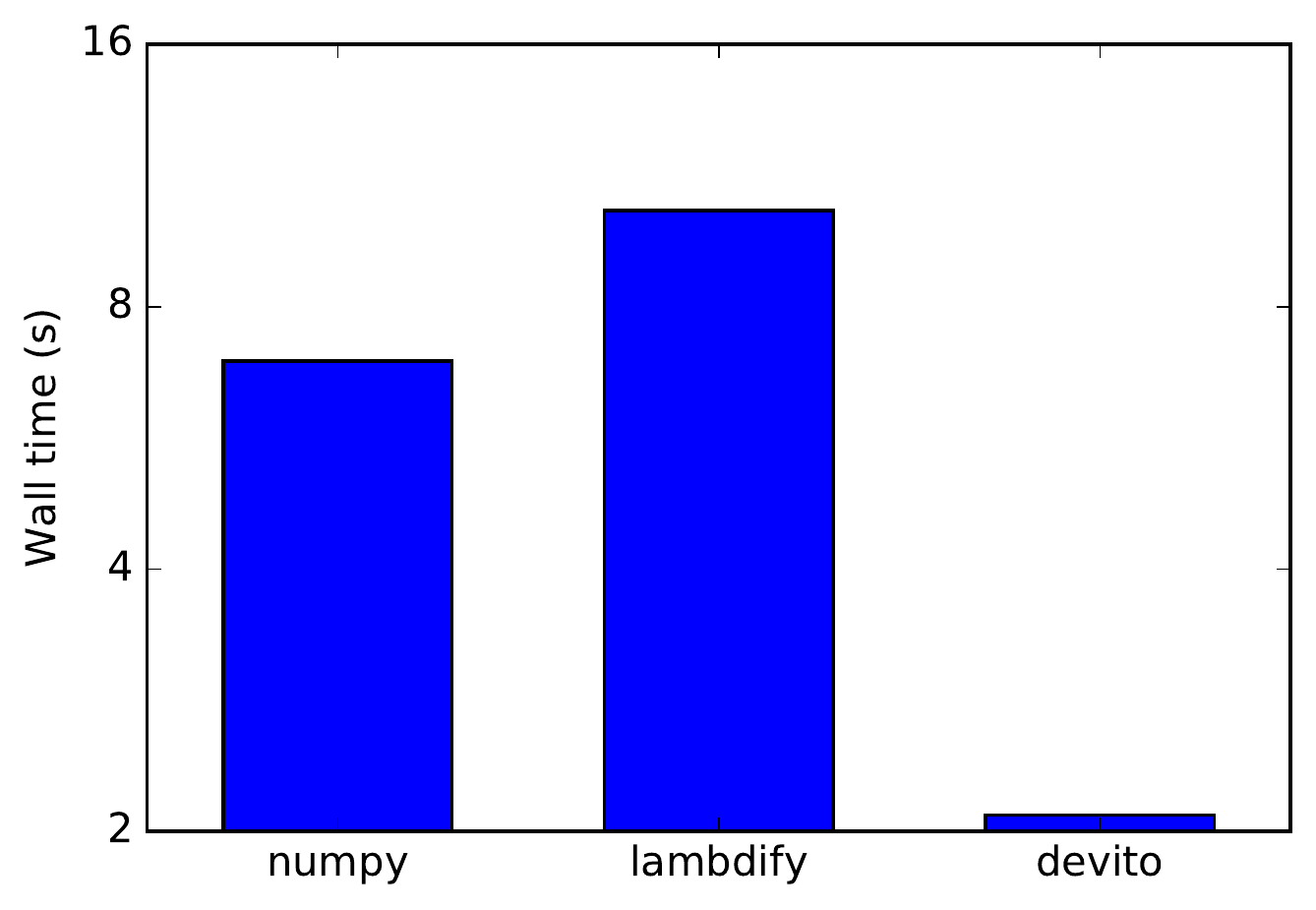}
\caption{Serial runtime comparison between \numpy{}, \sympy{} and
  Devito for the diffusion example.}
\label{fig:diffusion_performance}
\end{figure}

To demonstrate the performance advantages of custom code generation
and JIT compilation as performed in Devito over traditional approaches
to Python acceleration, we have compared the single core performance
of the vectorized \numpy{} and \sympy{} implementations of the diffusion
example. For this purpose the diffusion benchmark was run for 500
timesteps with a grid size of 0.001 on a single core of a \Intel Core
i3-4030U CPU. The results presented in
\figref{fig:diffusion_performance} show a clear performance advantage
of the DSL approach, as well as highlighting a significant overhead of
the vectorized \sympy{} implementation over the pure \numpy{}
approach. It is important to note here that the Devito DSL is the only
benchmarked approach that also encapsulates the timestepping loop
in the compilation step and that further performance advantages can be
expected when shared-memory parallelism is utilised, as we will show in
\secref{sec:acoustic_performance}.

\section{Seismic Inversion Example}

The key motivating application behind the creation of the Devito
finite fifference DSL are seismic exploration problems including
advanced modelling and adjoint-based inversion. These not only require
a fast and accurate wave propagation model, but also efficient,
easy-to-define and rigorous adjoint operators. In the following
example we demonstrate the symbolic definition of modelling and
adjoint modelling operations for the acoustic wave equation denoted
as:
\begin{equation} \label{eqn:acoustic}
  m \frac{\partial^2 u}{\partial t^2} + \eta \frac{\partial u}{\partial t} - \nabla^2 u = q
\end{equation}
where $u$ denotes the pressure wave field, $m$ is the square slowness,
$q$ is the source term and $\eta$ denotes the spatially varying
dampening factor used to implement an absorbing boundary condition.

\subsection{Acoustic wave equation}
\label{sec:acoustic}

The implementation of the forward operator with an explicit time marching
scheme is shown in \lstref{lst:adjoint}. First, the symbolic data
objects are created, where \lstinline!u! is defined as a time-varying
function, whereas \lstinline!m! and \lstinline[mathescape]!$\eta$! are
spatial grids only. Notably, the source term \lstinline!q! is missing
from the high-level PDE description, since the injection of the source
requires sparse point-to-grid interpolation. Since interpolation does
not fall within a finite difference abstraction, it is dealt with via
separate expressions, as further explained in
\secref{sec:interpolation}

It is also important to note here that the spatial order of the
derivative stencils in the operator definition is provided as an
argument and can thus be changed easily by the user, and that the
entire formulation is independent of problem dimension. From the
individual component symbols we create the wave equation in a purely
symbolic form, leaving Devito's \lstinline!dt!, \lstinline!dt2!  and
\lstinline!laplace! operators to perform the stencil expansion
according to the spatial and temporal discretization order. The
resulting expression is then rearranged using the \sympy{}
\lstinline!solve!  routine to denote forward propagation, from which
the corresponding \lstinline!Operator! can be created and executed.

\begin{lstlisting}[
    caption=Example code to solve the wave equation,
    label={lst:wave}, mathescape=true, float, floatplacement=HT]
def forward(model, nt, dt, h, order=2):
    shape = model.shape
    m = DenseData(name="m", shape=shape,
                  space_order=order)
    m.data[:] = model  # Set model data
    u = TimeData(name='u', shape=model.shape,
                 time_dim=nt, time_order=2,
                 space_order=order)
    $\eta$ = DenseData(name="$\eta$", shape=shape,
                  space_order=order)

    # Derive stencil from symbolic equation
    eqn = m * u.dt2 - u.laplace + $\eta$ * u.dt
    stencil = solve(eqn, u.forward)[0]

    op = Operator(stencils=Eq(u.forward, stencil),
                  nt=nt, subs={s: dt, h: h},
                  shape=shape, forward=True)
    # Source injection code omitted for brevity

    op.apply()
\end{lstlisting}

A very similar symbolic definition can be used to define the adjoint
operator for the wave equation, as demonstrated in Listing
\ref{lst:adjoint}. The key difference to the forward implementation
is that the dampening term \lstinline[mathescape]! $\eta$ * u.dt! is subtracted rather
than added, and that we rearrange the stencil expression for execution
backward in time. The shorthand notation \lstinline!u.forward! and
\lstinline!u.backward! hereby denote the highest and lowest stencil
point in the second-order time discretization stencil,
\lstinline!t + s! and \lstinline!t - s! respectively. These can be
used to ensure the correct time marching direction once the symbolic
function offsets \lstinline!s! and \lstinline!h! have been resolved to
explicit grid indices in the final stencil expression.

\begin{lstlisting}[
    caption=Example code to solve the adjoint wave equation,
    label={lst:adjoint}, mathescape=true, float, floatplacement=HT]
def adjoint(model, nt, dt, h, spc_order=2):
    shape = model.shape
    m = DenseData(name="m", shape=shape,
                  space_order=order)
    m.data[:] = model
    v = TimeData(name='v', shape=shape,
                 time_dim=nt, time_order=2,
                 space_order=spc_order)
    $\eta$ = DenseData(name="$\eta$", shape=shape,
                  space_order=order)

    # Derive stencil from symbolic equation
    eqn = m * v.dt2 - v.laplace - $\eta$ * v.dt
    stencil = solve(eqn, v.backward)[0]

    op = Operator(stencils=Eq(u.backward, stencil),
                  nt=nt, subs={s: dt, h: h},
                  shape=shape, forward=False)
    # Receiver interpolation omitted for brevity

    op.apply()
\end{lstlisting}

\subsection{Sparse point interpolation}
\label{sec:interpolation}

One key restriction of the high-level PDE notation used in the
previous examples is the inability to express interpolation between
irregularly spaced sparse points and the grid. This feature is
frequently required to inject source terms by
point-to-grid interpolation and sample the resulting wave field at
non-aligned sparse points via grid-to-point interpolation.

\begin{lstlisting}[
    caption=Example code for point-to-grid interpolation.,
    label={lst:point2grid}, float, floatplacement=HT]
eqns = [Eq(u[t, x, y], u[t, x, y] + expr),
        Eq(u[t, x+1, y], u[t, x+1, y] + expr),
        Eq(u[t, x, y+1], u[t, x, y+1] + expr),
        Eq(u[t, x+1, y+1], u[t, x+1, y+1] + expr)]
it = Iteration(eqns, index=p, limits=source.shape)]
\end{lstlisting}

\lstref{lst:point2grid} demonstrates how to construct a set of
expressions to perform point-to-grid interpolation, where each grid
value enclosing the sparse point is updated according to a source
term. Most notably, the update to the grid value in the wave field
\lstinline!u! is created using the low-level indexed variable format,
allowing us to define the explicit offsets needed to update every
enclosing grid point with the relevant source expression
\lstinline!expr!. A similar set of expressions can be generated to
perform the inverse interpolation from grid values to sparse points,
but an example of this is omitted here for brevity. Grid-to-point
interpolation, however, is of particular importance for seismic
inversion problems, since the behaviour of sparse points changes
between the forward and the adjoint runs.

Since there are potentially many sources in a wave field model, we can
also construct an additional loop around the expression set to iterate
over all sparse points contained in \lstinline!source!. For this,
Devito provides an additional low-level \lstinline!Iteration! class
that adds a user-defined loop construct into the auto-generated code.
This enables custom iterations over source source points and their
associated coordinate data, allowing users to create
coordinate-dependent symbolic expressions that define the interpolation
for each reference cell point.

\subsection{Validation}

We pointed out the requirement to have rigorous adjoint in order to
use our framework in seismic inversion. To validate our implementation
we use the simple definition of an adjoint as a test. For a given
matrix $A$, its adjoint is defined as the matrix $A^T$ such that:
\begin{eqnarray} \label{eqn:adjtest}
	&\forall (x, y),\nonumber\\
	&< Ax, y> - <x, A^Ty> = 0,\\
	&\frac{< Ax, y>}{<x, A^Ty>} = 1,\nonumber\\
\end{eqnarray}

where $<.,.>$ denotes the inner product. In the case of the wave
equation, $A$ represent the modelling operator while $A^T$ represents
the adjoint modelling operator, and from the definition in Equation
\ref{eqn:adjtest} we can compare the output of our kernels to verify
the accuracy of the adjoint. We validate our implementation with this
test in a two-dimensional and three-dimensional setting for multiple
spacial discretization order in Table \ref{adjres} .

\begin{table*}
\centering
\caption{Adjoint test for different discretization orders in 2D and
  3D. The test is computed on a two layer model.}
\label{adjres}
\begin{tabular}{|c|c|c||c|c|c|}
Order & Dimension & $<Ax,y>$ & $<x,A^Ty>$ &
           Difference & ratio \\
\hline
2nd order      &  2D:  & 373323.7976& 373323.7975434& 6.07169350e-05  &  1.0\\
4th order      &  2D:  & 340158.1486& 340158.1485252&  0.00012756     &  1.0\\
6th order      &  2D:  & 341557.3948& 341557.3947399&  0.00014287     &  1.0\\
8th order      &  2D:  & 358240.8513& 358240.8511931&  0.00016741     &  1.0\\
10th order     &  2D:  & 393488.5561& 393488.5559269&  0.00023841     &  1.0\\
12th order     &  2D:  & 439561.4005& 439561.4002034&  0.00035794     &  1.0\\
2nd order      &  3D:  & 2.17496552 & 2.174965534979&  -1.23030883e-08&  0.99999999\\
4th order      &  3D:  & 3.64447937 & 3.644479393901&  -2.13132316e-08&  0.99999999\\
6th order      &  3D:  & 3.78730372 & 3.787303745086&  -2.22477072e-08&  0.99999999\\
8th order      &  3D:  & 3.80286229 & 3.802862312852&  -2.23545817e-08&  0.99999999\\
10th order     &  3D:  & 3.80557957 & 3.805579596334&  -2.23736993e-08&  0.99999999\\
12th order     &  3D:  & 3.80318675 & 3.803186773144&  -2.23587757e-08&  0.99999999\\
\end{tabular}
\end{table*}

\subsection{Performance}
\label{sec:acoustic_performance}

\begin{figure}[ht] \centering
\begin{subfigure}{.5\textwidth} \centering
\includegraphics[width=\linewidth]{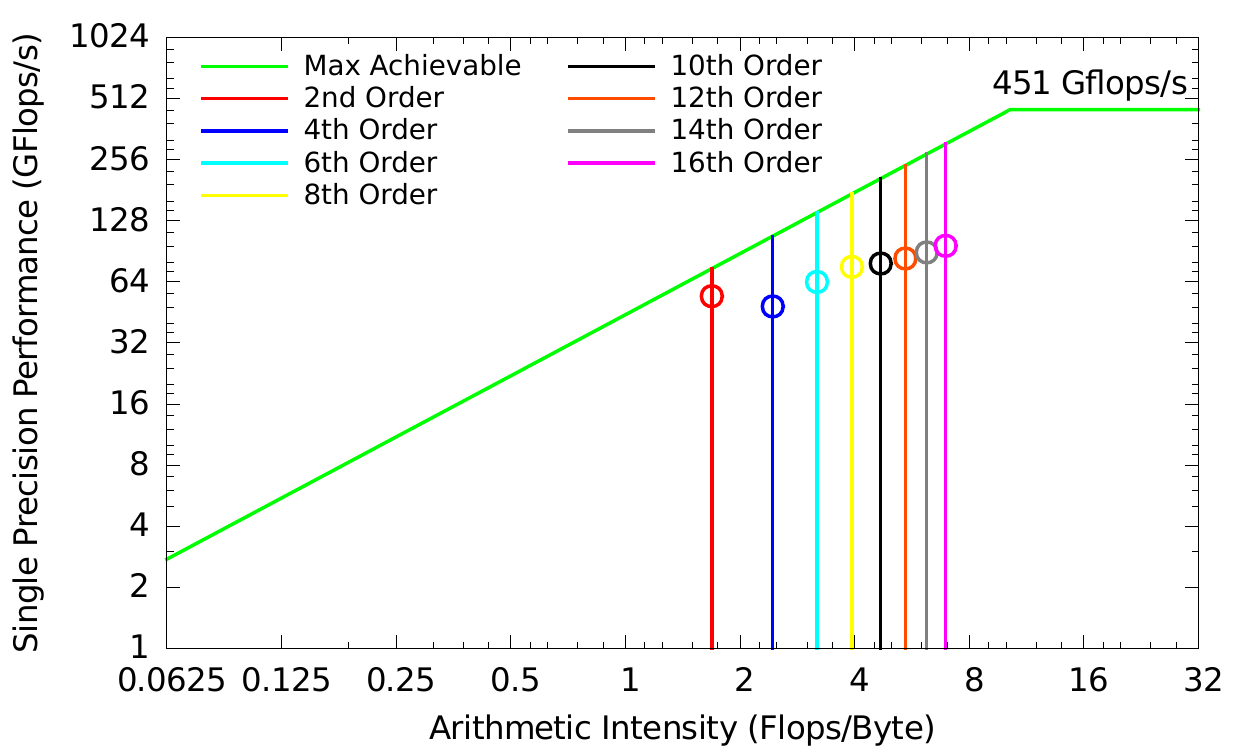}
\caption{\Xeon E5-2690v2 10C @ 3GHz}
\end{subfigure}

\begin{subfigure}{.5\textwidth} \centering
\includegraphics[width=\linewidth]{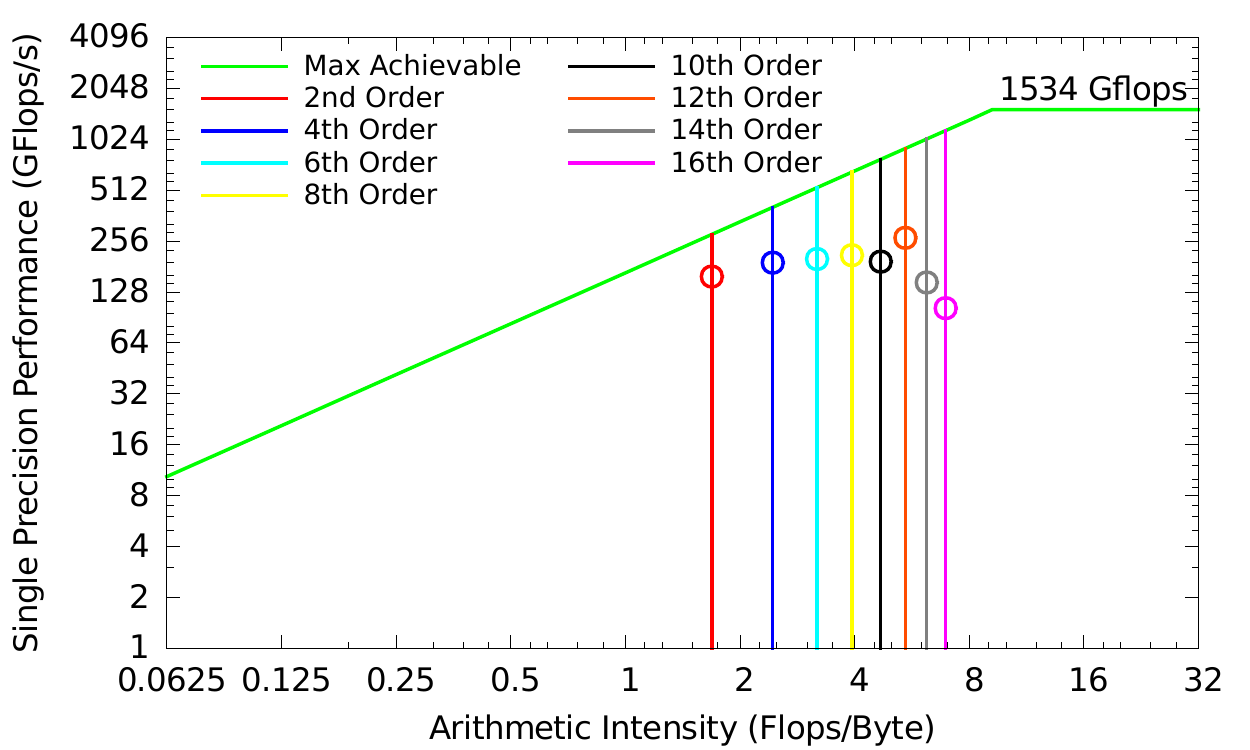}
\caption{\XeonPhi}
\end{subfigure}

\caption{Roofline plots for two target systems}
\label{fig:acoustic_performance}
\end{figure}

The performance of the finite difference code generated by Devito is
demonstrated on two target architectures, a \Xeon E5-2690v2
architecture with 10 physical cores running at 3GHz and a \XeonPhi
accelerator card, in \figref{fig:acoustic_performance}. The model used
for the benchmark is a gradient test with forward and adjoint
operators on a three-dimensional grid with $201 \times 201 \times 70$
grid points and $40$ PML grid points on each side, resulting in a
computational grid of size $281 \times 281 \times 150$. The grid size
is $15 m$ and the source term is a Ricker wavelet at $10 Hz$. The wave
field is modelled for 1 second with spatial discretizations of varying
order from 2 to 16. The results indicate that Devito is able to utilise
both architectures to a high degree of efficiency, while maintaining
the ability to increase accuracy by switching to higher order stencil
discretizations dynamically.

\section{Discussion}

In this paper we present the Devito finite difference DSL and showcase
its multi-level Python API to efficiently generate complex
mathematical operators from \sympy{} expressions. We highlight Devito's
practical applicability in creating forward and adjoint operators for
seismic inversion operators from concise symbolic definitions and
verify their correctness by showing that the generated code does in
fact behave mathematically like the adjoint of the equation. Moreover,
we demonstrate that the optimized code generated by Devito is capable
of achieving a significant percentage of peak performance on a
traditional CPU, as well as a \XeonPhi architecture.

When showcasing Devito's symbolic Python API we also compare it to
traditional approaches to accelerating Python computation and show
significant speed-ups on even a single core CPU. On top of the evident
performance benefits, a more concise formulation of simple
mathematical operators can be achieved using Devito, which increases
the productivity of domain scientists. Conversely, our total reliance
on the \sympy{} package demonstrates the importance of the Python
programming language and its vast software ecosystem in creating
performance-oriented DSLs and utilising them to build actual
scientific applications.

The use of Devito's second-layer API to create custom point-grid
interpolation routines further illustrates the complexities that often
prevent stencil DSLs from being used in practice. In the context of
seismic inversion, sparse point interpolation is the most important
feature that other finite difference DSLs lack. In addition, Devito 's
full compatibility with other scientific Python packages through the
\numpy{} array API entails that it is easy to implement a full-scale
inversion workflow in Python - including additional features such as
MPI parallelism between multiple inversion operators.

\section{Future Work}

Even though the primary target application for Devito is seismic
inversion, we aim to develop Devito towards becoming a generic finite
difference DSL. The diffusion example presented in
\secref{sec:diffusion_devito} shows that the current abstraction
generalises well to problems beyond the wave equation and its
additional low-level API provides an important degree of flexibility
when adding new features. Additional finite difference mechanisms,
such as staggered grids and new boundary condition types are of great
importance here to generalise Devito's applicability to other
application areas.

Additional improvements to the high and low-level symbolic APIs are
also planned that include customizable dimension symbols and the
automated derivation of iteration spaces. This additional automation
will in turn enable further integration with optimising compilers and
other external stencil frameworks and potentially even allow the
automated use of polyhedral compilation tools.

The seismic inversion operators demonstrated in this paper, and in
particular the sparse point interpolation features discussed in
\secref{sec:interpolation}, are still in an early development
stage. The seismic inversion example presented here uses the acoustic
isotropic wave equation, although, for real world applications, more
complex physical models involving anisotropy and elastic wave
propagation are commonly seen. In the future we plan to implement
these complex physical models using Devito, expanding Devito's
repertoire of features wherever necessary.

\section{Acknowledgements}

This work was financially supported in part by the Natural Sciences
and Engineering Research Council of Canada Collaborative Research and
Development Grant DNOISE II (CDRP J 375142-08), EPSRC grant
EP/L000407/1, the Imperial College London Intel Parallel Computing
Centre, SENAI CIMATEC and the MCTI (Federal Govt of Brazil)
scholarship MCTI/FINEP/CNPq 384214/2015-0. This research was carried
out as part of the SINBAD II project with the support of the member
organizations of the SINBAD
Consortium.

\bibliographystyle{IEEEtran}
\bibliography{IEEEabrv,references}

\begin{thebibliography}{10}
\providecommand{\url}[1]{#1}
\csname url@samestyle\endcsname
\providecommand{\newblock}{\relax}
\providecommand{\bibinfo}[2]{#2}
\providecommand{\BIBentrySTDinterwordspacing}{\spaceskip=0pt\relax}
\providecommand{\BIBentryALTinterwordstretchfactor}{4}
\providecommand{\BIBentryALTinterwordspacing}{\spaceskip=\fontdimen2\font plus
\BIBentryALTinterwordstretchfactor\fontdimen3\font minus
  \fontdimen4\font\relax}
\providecommand{\BIBforeignlanguage}[2]{{%
\expandafter\ifx\csname l@#1\endcsname\relax
\typeout{** WARNING: IEEEtran.bst: No hyphenation pattern has been}%
\typeout{** loaded for the language `#1'. Using the pattern for}%
\typeout{** the default language instead.}%
\else
\language=\csname l@#1\endcsname
\fi
#2}}
\providecommand{\BIBdecl}{\relax}
\BIBdecl

\bibitem{Oliphant2007}
T.~E. Oliphant, ``Python for scientific computing,'' \emph{Computing in Science
  Engineering}, vol.~9, no.~3, pp. 10--20, May 2007.

\bibitem{Millman2011}
K.~J. Millman and M.~Aivazis, ``Python for scientists and engineers,''
  \emph{Computing in Science Engineering}, vol.~13, no.~2, pp. 9--12, March
  2011.

\bibitem{Perez2011}
F.~Perez, B.~E. Granger, and J.~D. Hunter, ``Python: An ecosystem for
  scientific computing,'' \emph{Computing in Science Engineering}, vol.~13,
  no.~2, pp. 13--21, March 2011.

\bibitem{Behnel2011}
S.~Behnel, R.~Bradshaw, C.~Citro, L.~Dalcin, D.~S. Seljebotn, and K.~Smith,
  ``Cython: The best of both worlds,'' \emph{Computing in Science Engineering},
  vol.~13, no.~2, pp. 31--39, March 2011.

\bibitem{Lam2015}
\BIBentryALTinterwordspacing
S.~K. Lam, A.~Pitrou, and S.~Seibert, ``Numba: A llvm-based python jit
  compiler,'' in \emph{Proceedings of the Second Workshop on the LLVM Compiler
  Infrastructure in HPC}, ser. LLVM '15.\hskip 1em plus 0.5em minus 0.4em\relax
  New York, NY, USA: ACM, 2015, pp. 7:1--7:6. [Online]. Available:
  \url{http://doi.acm.org/10.1145/2833157.2833162}
\BIBentrySTDinterwordspacing

\bibitem{Logg2012}
A.~Logg, K.-A. Mardal, and G.~Wells, \emph{Automated Solution of Differential
  Equations by the Finite Element Method: The FEniCS Book}.\hskip 1em plus
  0.5em minus 0.4em\relax Springer Publishing Company, Incorporated, 2012.

\bibitem{Meuer2016}
A.~Meurer, C.~P. Smith, M.~Paprocki, O.~{\v{C}}ert{\'\i}k, M.~Rocklin,
  A.~Kumar, S.~Ivanov, J.~K. Moore, S.~Singh, T.~Rathnayake \emph{et~al.},
  ``Sympy: Symbolic computing in python,'' PeerJ Preprints, Tech. Rep., 2016.

\bibitem{vanDerWalt2011}
S.~van~der Walt, S.~C. Colbert, and G.~Varoquaux, ``The numpy array: A
  structure for efficient numerical computation,'' \emph{Computing in Science
  Engineering}, vol.~13, no.~2, pp. 22--30, March 2011.

\bibitem{Cardenas1970}
A.~F. C{\'a}rdenas and W.~J. Karplus, ``Pdel—a language for partial
  differential equations,'' \emph{Communications of the ACM}, vol.~13, no.~3,
  pp. 184--191, 1970.

\bibitem{Grant1988}
G.~O. Cook~Jr, ``Alpal: A tool for the development of large-scale simulation
  codes,'' Lawrence Livermore National Lab., CA (USA), Tech. Rep., 1988.

\bibitem{Yukio1985}
Y.~Umetani, ``Deqsol a numerical simulation language for vector/parallel
  processors,'' \emph{Proc. IFIP TC2/WG22, 1985}, vol.~5, pp. 147--164, 1985.

\bibitem{VanEngelen1996}
R.~Van~Engelen, L.~Wolters, and G.~Cats, ``Ctadel: A generator of
  multi-platform high performance codes for pde-based scientific
  applications,'' in \emph{Proceedings of the 10th international conference on
  Supercomputing}.\hskip 1em plus 0.5em minus 0.4em\relax ACM, 1996, pp.
  86--93.

\bibitem{Rathgeber2015}
F.~Rathgeber, D.~A. Ham, L.~Mitchell, M.~Lange, F.~Luporini, A.~T. McRae, G.-T.
  Bercea, G.~R. Markall, and P.~H. Kelly, ``Firedrake: automating the finite
  element method by composing abstractions,'' \emph{Submitted to ACM TOMS},
  2015.

\bibitem{Alnaes2014}
M.~S. Aln{\ae}s, A.~Logg, K.~B. {\O}lgaard, M.~E. Rognes, and G.~N. Wells,
  ``{U}nified {F}orm {L}anguage: a domain-specific language for weak
  formulations of partial differential equations,'' \emph{ACM Transactions on
  Mathematical Software (TOMS)}, vol.~40, no.~2, p.~9, 2014.

\bibitem{Farrell2013}
\BIBentryALTinterwordspacing
P.~E. Farrell, D.~A. Ham, S.~W. Funke, and M.~E. Rognes, ``Automated derivation
  of the adjoint of high-level transient finite element programs,'' \emph{SIAM
  Journal on Scientific Computing}, vol.~35, no.~4, pp. C369--C393, 2013.
  [Online]. Available: \url{http://dx.doi.org/10.1137/120873558}
\BIBentrySTDinterwordspacing

\bibitem{Brandvik2010}
\BIBentryALTinterwordspacing
T.~Brandvik and G.~Pullan, ``Sblock: A framework for efficient stencil-based
  pde solvers on multi-core platforms,'' in \emph{Proceedings of the 2010 10th
  IEEE International Conference on Computer and Information Technology}, ser.
  CIT '10.\hskip 1em plus 0.5em minus 0.4em\relax Washington, DC, USA: IEEE
  Computer Society, 2010, pp. 1181--1188. [Online]. Available:
  \url{http://dx.doi.org/10.1109/CIT.2010.214}
\BIBentrySTDinterwordspacing

\bibitem{Henretty2013}
\BIBentryALTinterwordspacing
T.~Henretty, R.~Veras, F.~Franchetti, L.-N. Pouchet, J.~Ramanujam, and
  P.~Sadayappan, ``A stencil compiler for short-vector simd architectures,'' in
  \emph{Proceedings of the 27th International ACM Conference on International
  Conference on Supercomputing}, ser. ICS '13.\hskip 1em plus 0.5em minus
  0.4em\relax New York, NY, USA: ACM, 2013, pp. 13--24. [Online]. Available:
  \url{http://doi.acm.org/10.1145/2464996.2467268}
\BIBentrySTDinterwordspacing

\bibitem{Zhang2012}
\BIBentryALTinterwordspacing
Y.~Zhang and F.~Mueller, ``Auto-generation and auto-tuning of 3d stencil codes
  on gpu clusters,'' in \emph{Proceedings of the Tenth International Symposium
  on Code Generation and Optimization}, ser. CGO '12.\hskip 1em plus 0.5em
  minus 0.4em\relax New York, NY, USA: ACM, 2012, pp. 155--164. [Online].
  Available: \url{http://doi.acm.org/10.1145/2259016.2259037}
\BIBentrySTDinterwordspacing

\bibitem{Datta2008}
\BIBentryALTinterwordspacing
K.~Datta, M.~Murphy, V.~Volkov, S.~Williams, J.~Carter, L.~Oliker,
  D.~Patterson, J.~Shalf, and K.~Yelick, ``Stencil computation optimization and
  auto-tuning on state-of-the-art multicore architectures,'' in
  \emph{Proceedings of the 2008 ACM/IEEE Conference on Supercomputing}, ser. SC
  '08.\hskip 1em plus 0.5em minus 0.4em\relax Piscataway, NJ, USA: IEEE Press,
  2008, pp. 4:1--4:12. [Online]. Available:
  \url{http://dl.acm.org/citation.cfm?id=1413370.1413375}
\BIBentrySTDinterwordspacing

\bibitem{PATUS}
\BIBentryALTinterwordspacing
M.~Christen, O.~Schenk, and H.~Burkhart, ``Patus: A code generation and
  autotuning framework for parallel iterative stencil computations on modern
  microarchitectures,'' in \emph{Proceedings of the 2011 IEEE International
  Parallel \& Distributed Processing Symposium}, ser. IPDPS '11.\hskip 1em plus
  0.5em minus 0.4em\relax Washington, DC, USA: IEEE Computer Society, 2011, pp.
  676--687. [Online]. Available: \url{http://dx.doi.org/10.1109/IPDPS.2011.70}
\BIBentrySTDinterwordspacing

\bibitem{STARGATES}
K.~A. Hawick and D.~P. Playne, ``Simulation software generation using a
  domain-specific language for partial differential field equations,'' in
  \emph{11th International Conference on Software Engineering Research and
  Practice (SERP’13)}, no. CSTN-187.\hskip 1em plus 0.5em minus 0.4em\relax
  Las Vegas, USA: WorldComp, 22-25 July 2013, p. SER3829.

\bibitem{HiPACC}
R.~Membarth, F.~Hannig, J.~Teich, and H.~K{\"o}stler, ``Towards domain-specific
  computing for stencil codes in hpc,'' in \emph{High Performance Computing,
  Networking, Storage and Analysis (SCC), 2012 SC Companion:}.\hskip 1em plus
  0.5em minus 0.4em\relax IEEE, 2012, pp. 1133--1138.

\bibitem{Mudalidge2012}
G.~R. Mudalige, M.~B. Giles, I.~Reguly, C.~Bertolli, and P.~H.~J. Kelly, ``Op2:
  An active library framework for solving unstructured mesh-based applications
  on multi-core and many-core architectures,'' in \emph{Innovative Parallel
  Computing (InPar), 2012}, May 2012, pp. 1--12.

\bibitem{Reguly2014}
\BIBentryALTinterwordspacing
I.~Z. Reguly, G.~R. Mudalige, M.~B. Giles, D.~Curran, and S.~McIntosh-Smith,
  ``The ops domain specific abstraction for multi-block structured grid
  computations,'' in \emph{Proceedings of the Fourth International Workshop on
  Domain-Specific Languages and High-Level Frameworks for High Performance
  Computing}, ser. WOLFHPC '14.\hskip 1em plus 0.5em minus 0.4em\relax
  Piscataway, NJ, USA: IEEE Press, 2014, pp. 58--67. [Online]. Available:
  \url{http://dx.doi.org/10.1109/WOLFHPC.2014.7}
\BIBentrySTDinterwordspacing

\bibitem{Howes2008}
\BIBentryALTinterwordspacing
L.~W. Howes, A.~Lokhmotov, A.~F. Donaldson, and P.~H. Kelly, ``Deriving
  efficient data movement from decoupled access/execute specifications,'' in
  \emph{Proceedings of the 4th International Conference on High Performance
  Embedded Architectures and Compilers}, ser. HiPEAC '09.\hskip 1em plus 0.5em
  minus 0.4em\relax Berlin, Heidelberg: Springer-Verlag, 2009, pp. 168--182.
  [Online]. Available: \url{http://dx.doi.org/10.1007/978-3-540-92990-1\_14}
\BIBentrySTDinterwordspacing

\bibitem{Rathgeber2012}
F.~Rathgeber, G.~R. Markall, L.~Mitchell, N.~Loriant, D.~A. Ham, C.~Bertolli,
  and P.~H. Kelly, ``Pyop2: A high-level framework for performance-portable
  simulations on unstructured meshes,'' in \emph{High Performance Computing,
  Networking Storage and Analysis, SC Companion:}.\hskip 1em plus 0.5em minus
  0.4em\relax Los Alamitos, CA, USA: IEEE Computer Society, 2012, pp.
  1116--1123.

\bibitem{Halide}
\BIBentryALTinterwordspacing
J.~Ragan-Kelley, C.~Barnes, A.~Adams, S.~Paris, F.~Durand, and S.~Amarasinghe,
  ``Halide: A language and compiler for optimizing parallelism, locality, and
  recomputation in image processing pipelines,'' in \emph{Proceedings of the
  34th ACM SIGPLAN Conference on Programming Language Design and
  Implementation}, ser. PLDI '13.\hskip 1em plus 0.5em minus 0.4em\relax New
  York, NY, USA: ACM, 2013, pp. 519--530. [Online]. Available:
  \url{http://doi.acm.org/10.1145/2491956.2462176}
\BIBentrySTDinterwordspacing

\end{thebibliography}

\end{document}